\documentstyle[twoside]{article}

\catcode`\@=11
\long\def\@makefntext#1{
\protect\noindent \hbox to 3.2pt {\hskip-.9pt  
$^{{\eightrm\@thefnmark}}$\hfil}#1\hfill}		%CAN BE USED 

\def\thefootnote{\fnsymbol{footnote}}
\def\@makefnmark{\hbox to 0pt{$^{\@thefnmark}$\hss}}	%ORIGINAL 
	
\def\ps@myheadings{\let\@mkboth\@gobbletwo
\def\@oddhead{\hbox{}
\rightmark\hfil\eightrm\thepage}   
\def\@oddfoot{}\def\@evenhead{\eightrm\thepage\hfil
\leftmark\hbox{}}\def\@evenfoot{}
\def\sectionmark##1{}\def\subsectionmark##1{}}

\oddsidemargin=\evensidemargin
\renewcommand{\thefootnote}{\fnsymbol{footnote}}

\newcounter{sectionc}\newcounter{subsectionc}\newcounter{subsubsectionc}
\renewcommand{\section}[1] {\vspace{12pt}\addtocounter{sectionc}{1} 
\setcounter{subsectionc}{0}\setcounter{subsubsectionc}{0}\noindent 
	{\tenbf\thesectionc. #1}\par\vspace{5pt}}
\renewcommand{\subsection}[1] {\vspace{12pt}\addtocounter{subsectionc}{1} 
	\setcounter{subsubsectionc}{0}\noindent 
	{\bf\thesectionc.\thesubsectionc. {\kern1pt \bfit #1}}\par\vspace{5pt}}
\renewcommand{\subsubsection}[1] {\vspace{12pt}\addtocounter{subsubsectionc}{1}
	\noindent{\tenrm\thesectionc.\thesubsectionc.\thesubsubsectionc.
	{\kern1pt \tenit #1}}\par\vspace{5pt}}
\newcommand{\nonumsection}[1] {\vspace{12pt}\noindent{\tenbf #1}
	\par\vspace{5pt}}

\newcounter{appendixc}
\newcounter{subappendixc}[appendixc]
\newcounter{subsubappendixc}[subappendixc]
\renewcommand{\thesubappendixc}{\Alph{appendixc}.\arabic{subappendixc}}
\renewcommand{\thesubsubappendixc}
	{\Alph{appendixc}.\arabic{subappendixc}.\arabic{subsubappendixc}}

\renewcommand{\appendix}[1] {\vspace{12pt}
        \refstepcounter{appendixc}
        \setcounter{figure}{0}
        \setcounter{table}{0}
        \setcounter{lemma}{0}
        \setcounter{theorem}{0}
        \setcounter{corollary}{0}
        \setcounter{definition}{0}
        \setcounter{equation}{0}
        \renewcommand{\thefigure}{\Alph{appendixc}.\arabic{figure}}
        \renewcommand{\thetable}{\Alph{appendixc}.\arabic{table}}
        \renewcommand{\theappendixc}{\Alph{appendixc}}
        \renewcommand{\thelemma}{\Alph{appendixc}.\arabic{lemma}}
        \renewcommand{\thetheorem}{\Alph{appendixc}.\arabic{theorem}}
        \renewcommand{\thedefinition}{\Alph{appendixc}.\arabic{definition}}
        \renewcommand{\thecorollary}{\Alph{appendixc}.\arabic{corollary}}
        \renewcommand{\theequation}{\Alph{appendixc}.\arabic{equation}}
        \noindent{\tenbf Appendix \theappendixc #1}\par\vspace{5pt}}
\newcommand{\subappendix}[1] {\vspace{12pt}
        \refstepcounter{subappendixc}
        \noindent{\bf Appendix \thesubappendixc. {\kern1pt \bfit #1}}
	\par\vspace{5pt}}
\newcommand{\subsubappendix}[1] {\vspace{12pt}
        \refstepcounter{subsubappendixc}
        \noindent{\rm Appendix \thesubsubappendixc. {\kern1pt \tenit #1}}
	\par\vspace{5pt}}

\newcommand{\textlineskip}{\baselineskip=13pt}
\newcommand{\smalllineskip}{\baselineskip=10pt}

\def\eightcirc{
\begin{picture}(0,0)
\put(4.4,1.8){\circle{6.5}}
\end{picture}}
\def\eightcopyright{\eightcirc\kern2.7pt\hbox{\eightrm c}} 

\newcommand{\copyrightheading}[1]
	{\vspace*{-2.5cm}\smalllineskip{\flushright
	{PSU-TH/234}\\
	{November 2000}\\
	 }}
\def\abstracts#1#2#3{{
	\centering{\begin{minipage}{4.5in}\baselineskip=10pt\footnotesize
	\parindent=0pt #1\par 
	\parindent=15pt #2\par
	\parindent=15pt #3
	\end{minipage}}\par}} 

\renewenvironment{thebibliography}[1]
	{\frenchspacing
	 \ninerm\baselineskip=11pt
	 \begin{list}{\arabic{enumi}.}
	{\usecounter{enumi}\setlength{\parsep}{0pt}
	 \setlength{\leftmargin 12.7pt}{\rightmargin 0pt} %FOR 1--9 ITEMS
	 \setlength{\itemsep}{0pt} \settowidth
	{\labelwidth}{#1.}\sloppy}}{\end{list}}

\newcounter{itemlistc}
\newcounter{romanlistc}
\newcounter{alphlistc}
\newcounter{arabiclistc}

\newcommand{\fcaption}[1]{
        \refstepcounter{figure}
        \setbox\@tempboxa = \hbox{\footnotesize Fig.~\thefigure. #1}
        \ifdim \wd\@tempboxa > 5in
           {\begin{center}
        \parbox{5in}{\footnotesize\smalllineskip Fig.~\thefigure. #1}
            \end{center}}
        \else
             {\begin{center}
             {\footnotesize Fig.~\thefigure. #1}
              \end{center}}
        \fi}

\newcommand{\tcaption}[1]{
        \refstepcounter{table}
        \setbox\@tempboxa = \hbox{\footnotesize Table~\thetable. #1}
        \ifdim \wd\@tempboxa > 5in
           {\begin{center}
        \parbox{5in}{\footnotesize\smalllineskip Table~\thetable. #1}
            \end{center}}
        \else
             {\begin{center}
             {\footnotesize Table~\thetable. #1}
              \end{center}}
        \fi}

\def\@citex[#1]#2{\if@filesw\immediate\write\@auxout
	{\string\citation{#2}}\fi
\def\@citea{}\@cite{\@for\@citeb:=#2\do
	{\@citea\def\@citea{,}\@ifundefined
	{b@\@citeb}{{\bf ?}\@warning
	{Citation `\@citeb' on page \thepage \space undefined}}
	{\csname b@\@citeb\endcsname}}}{#1}}

\newif\if@cghi
\def\cite{\@cghitrue\@ifnextchar [{\@tempswatrue
	\@citex}{\@tempswafalse\@citex[]}}
\def\citelow{\@cghifalse\@ifnextchar [{\@tempswatrue
	\@citex}{\@tempswafalse\@citex[]}}
\def\@cite#1#2{{$\null^{#1}$\if@tempswa\typeout
	{IJCGA warning: optional citation argument 
	ignored: `#2'} \fi}}

\def\pmb#1{\setbox0=\hbox{#1}
	\kern-.025em\copy0\kern-\wd0
	\kern.05em\copy0\kern-\wd0
	\kern-.025em\raise.0433em\box0}

\def\fnt#1#2{\footnotetext{\kern-.3em
	{$^{\mbox{\scriptsize #1}}$}{#2}}}

\def\fpage#1{\begingroup
\voffset=.3in
\thispagestyle{empty}\begin{table}[b]\centerline{\footnotesize #1}
	\end{table}\endgroup}

\font\tenrm=cmr10
\font\tenit=cmti10 
\font\tenbf=cmbx10
\font\bfit=cmbxti10 at 10pt
\font\ninerm=cmr9

\font\eightrm=cmr8

\def\qed{\hbox{${\vcenter{\vbox{			%HOLLOW SQUARE
   \hrule height 0.4pt\hbox{\vrule width 0.4pt height 6pt
   \kern5pt\vrule width 0.4pt}\hrule height 0.4pt}}}$}}

\renewcommand{\thefootnote}{\fnsymbol{footnote}}	%USE SYMBOLIC FOOTNOTE

\input epsf.tex
\def\desepsf(#1 width #2){\epsfxsize=#2 \epsfbox{#1}}
\begin{document}

%\runninghead{Parton showers beyond the leading order 
%$\ldots$} {Parton showers beyond the leading order 
%$\ldots$}

\normalsize\textlineskip
\setcounter{page}{1}

\copyrightheading{}			%{Vol. 0, No. 0 (1993) 000--000}

\vspace*{0.88truein}

\fpage{1}
\centerline{\bf PARTON SHOWERS BEYOND THE LEADING ORDER: }
\vspace*{0.035truein}
\centerline{\bf A FACTORIZATION APPROACH
\footnote{Talk at the 
DPF2000 Meeting, Ohio State University, 9-12 August 2000.}}
\vspace*{0.37truein}
\centerline{\footnotesize F. HAUTMANN}
\vspace*{0.035truein}
\centerline{\footnotesize\it  Department of Physics, Pennsylvania 
State University, University Park PA 16802}
\vspace*{10pt}

\vspace*{0.37truein}
\abstracts{
 We discuss recent work on methods for incorporating  
nonleading QCD  corrections in parton shower algorithms.   
}{}{}

\textlineskip			%) USE THIS MEASUREMENT WHEN THERE IS
\vspace*{12pt}			%) NO SECTION HEADING

%\vspace*{1pt}\textlineskip	%) USE THIS MEASUREMENT WHEN THERE IS
%\section{General Appearance}	%) A SECTION HEADING
%\vspace*{-0.5pt}
%\noindent

\textheight=7.8truein
\setcounter{footnote}{0}
\renewcommand{\thefootnote}{\alph{footnote}}

\vspace*{19pt}

 Parton shower 
Monte Carlo event generators are 
the main practical  tool to describe multi-particle final states in 
high energy collisions. 
 These event generators couple a 
leading-order (LO) hard scattering to a showering, treated basically   
 in the leading logarithm approximation.  
  Many nonleading effects are also included in these  
 event generators:
 for instance, through the exact multi-parton kinematics; 
through the angular ordering of 
gluon emission; through optimal choices of the 
renormalization scale in the running coupling.  In some cases 
 there are also treatments 
(see, e.g., refs.\cite{seymour,corcella,andre,mrenna})  
 to approximately include   next-to-leading-order 
(NLO)  corrections to the hard scattering. Nevertheless,  
 there is as yet 
no  method for going beyond the leading approximation systematically. 
This implies that event generators cannot 
  incorporate  fully  
   the known NLO (and NNLO)  calculations of 
hard-scattering cross sections. 

Recent  papers\cite{friberg,jccmonte,mcsept} 
 have  started to investigate systematic subtractive methods 
as a way to 
tackle this problem.  
This talk describes some of the work in this direction. 

To discuss this, let us   
  schematically represent the cross section in an 
  event generator as 
\begin{equation} 
\label{sigmaW}
   \sigma[W] = \sum_{{\rm final~states~} X} W(X)~
               {\rm{PS}} \otimes  \hat H .
\end{equation}
Here 
$W$ is a weight function  that specifies the 
definition of the particular cross section under consideration. 
The symbol ${\rm{PS}}$ denotes the parton shower and the symbol $\otimes$
denotes its action on the initial and final partons in the hard
scattering, whose cross section is denoted by $\hat H$.

In a standard  Monte Carlo, 
the hard scattering is taken to the 
 leading order, $H^{\rm{(LO)}}$, and ${\rm{PS}}$ denotes showering from 
the partons in $H^{\rm{(LO)}}$.  In an   NLO Monte Carlo,  
the cross section involves a structure of the form 
\begin{equation} 
\label{schemat}
   {\rm{PS}} \otimes  
   \left[ H^{\rm{(LO)}} + 
          \alpha_s  
          \left( H^{\rm{(NLO)}} 
               - {\rm{PS}}_I (1) \otimes H^{\rm(LO)} 
               - {\rm{PS}}_F (1) \otimes H^{\rm(LO)} 
          \right) 
  \right] .  
\end{equation} 
        Here the first term in the square brackets is the LO hard-scattering 
        function, 
        and the second term is the  
        subtracted NLO hard-scattering function.   
        $H^{\rm(NLO)}$ is 
        the result of computing the  partonic cross section from
        the NLO graphs, while    
        ${\rm{PS}}_I (1)$ and ${\rm{PS}}_F (1)$ are the order $\alpha_s$ 
        approximations to the initial-state and final-state showering. The 
        subtraction terms avoid double counting of events already included 
        by showering from $H^{\rm{(LO)}}$.

Ref.\cite{jccmonte} gives 
a simple example of the structure (\ref{schemat}): it treats 
the  photon-gluon fusion process 
in deep inelastic $e p$ scattering,   
using  the  Monte Carlo  algorithm 
 of ref.\cite{bengtsson}. 
This case is  special because   i) only 
 the subtraction term in 
${\rm{PS}}_I (1)$, coming from the initial state, is present, 
and ii) there are no leading-power contributions from soft gluons.   
Then the geometry of the 
leading-power regions 
that contribute to (\ref{schemat}) 
is simple, just   consisting of the ultraviolet region  
and the region collinear to the 
initial state, with no overlap.

For general NLO  processes,   
leading-power contributions to  (\ref{schemat})  involve  
 a complicated geometry of possibly
overlapping  regions, that include 
soft contributions as well as 
collinear contributions. A typical case is shown in    
Fig.~1.    
Ref.\cite{mcsept}  discusses how to extend the method to deal with 
such cases. A first step in this program  is  
to develop techniques to decompose Feynman graphs into sums of terms 
over   different  regions, with    
  the terms  arranged so as  to correspond to factors
  in a factorization formula  suitable for the 
  event-generator   application. 
Similar problems are encountered in  multi-loop 
calculations based on graph-by-graph methods (see, e.g., 
ref.\cite{binohein}). 
  This corresponds to a decomposition for $H^{\rm{(NLO)}}$ of the kind   
\begin{equation}
\label{decomp}
H^{\rm{(NLO)}} =  \sum_{{\rm{regions}} \ R} A_H (R) \ 
+ \;\; \mbox{nonleading power} ,                   
\end{equation}
 holding uniformly over the whole of the phase space. 
        Each of the pieces in 
         (\ref{decomp})  contains    
        counterterms that prevent double counting and provide the 
        suppression for going outside the region in which that particular 
        piece was originally supposed to give a good 
        approximation to the matrix 
        element. 
        This subtractive approach is to be contrasted with 
        approaches based on  splitting the 
        phase space in different domains and using different 
        approximations to 
        the matrix element in these different domains (see, e.g., 
        refs.\cite{seymour,corcella}).

\begin{figure}
\centerline{\desepsf(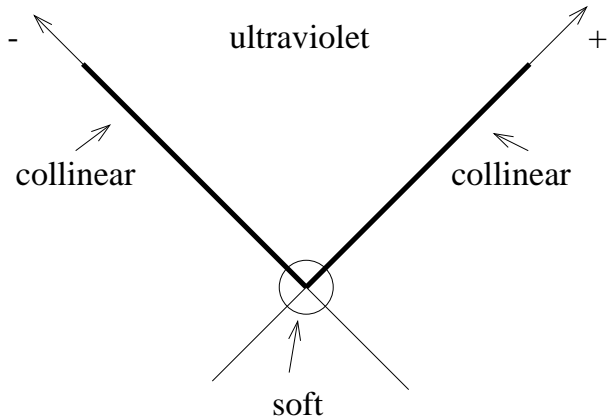 width 8 cm)}
\vspace*{5mm}
\caption{Geometry of leading-power regions 
for a typical  NLO cross section. The axes 
denote lightcone directions in momentum space. }
\label{fig:allreg}
\end{figure}

Semi-analytical\cite{catsey,frix} or fully 
numerical\cite{sopnum} subtraction methods have been devised to 
calculate NLO
quantities that are  infrared safe. These methods are not 
directly applicable in   event generators  
that simulate  
the fully exclusive structure  of the hadronic final states, 
 since here the  quantities being computed  are not 
infrared safe in perturbation theory. 
In particular, one cannot use a cancellation of soft gluon 
contributions between real and virtual graphs. 

Ref.\cite{nonlight} discusses a strategy to construct 
a decomposition of the kind ({\ref{decomp}). This is 
inspired by the R-operation techniques of renormalization.  
 See  ref.\cite{tka} for a related approach.  
Given the list of the leading  regions, 
determined by standard power counting arguments\cite{libby}, 
we proceed from ``smaller'' to ``larger'' regions. See Fig.~2. 
For each region $R$, we remove the contribution from smaller regions, 
and construct an approximation to the  matrix element valid in $R$, 
 up to power suppressed corrections. Then we  subtract 
any divergences that appear in this expression coming from 
larger regions. 
To ensure that 
the splitting between the terms is defined gauge-invariantly, 
at each step we demand that the 
counterterms be constructed from matrix elements of Wilson line 
operators, 
\begin{equation}
\label{VIVF}
V ( n ) = {\cal P}\exp\left(
  i g \int^{+\infty}_0 dy \, n \cdot A (y \, n ) 
  \right) ,    
\end{equation}
with suitable directions $n$ for the  lines. 
Evolution equations in  $n$ enable one to   
connect the results corresponding to different 
directions. 
Taking $n$ along lightlike directions gives back the 
eikonals of the leading infrared  approximation, while 
 taking $n$ off the light cone  provides effective cut-off 
parameters --- but in such a way that even a formalism involving off-shell
partons (see, e.g., ref.\cite{mrenna}) 
could be treated with gauge invariance.
It has been shown how to apply this procedure   
 to virtual loops\cite{nonlight}  
and real emission corrections\cite{mcsept}.

\begin{figure}
\centerline{\desepsf(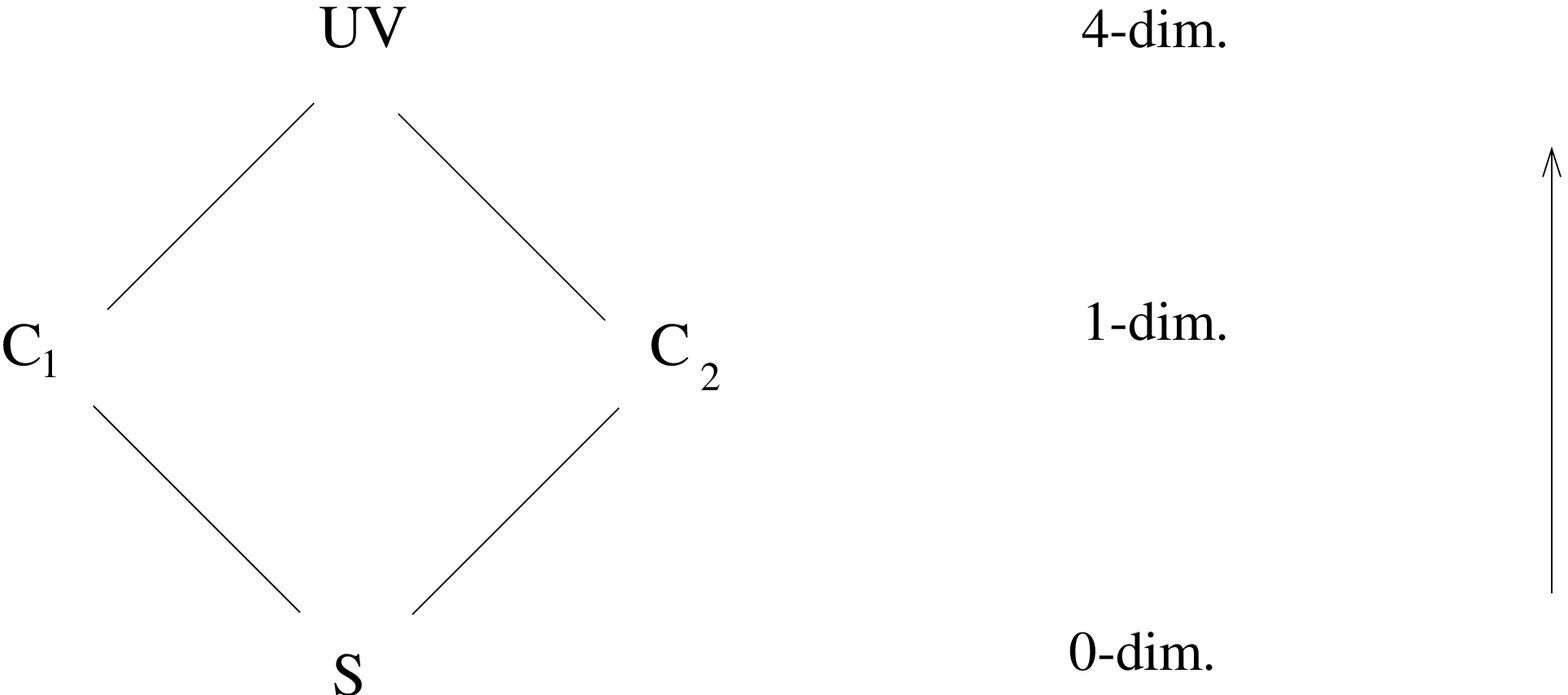 width 11 cm)}
\vspace*{5mm}
\caption{The soft (S), collinear (C$_1$, C$_2$), ultraviolet 
(UV) regions, corresponding to the model geometry of Fig.~1.  }
\label{fig:catalog}
\end{figure}

Ref.\cite{mcsept} uses this method to construct a decomposition 
 (\ref{decomp}) for  one-gluon emission graphs in 
deep inelastic scattering. 
         The 
        term
        corresponding to the ultraviolet region 
        gives  the 
        subtracted hard-scattering function to be used in
        Eq.~(\ref{schemat}). The collinear terms  correspond to the
        evolution kernels to be used in the showering.  The 
        soft term would correspond to a new element in the
        Monte Carlo, but  
         ref.\cite{mcsept} shows   
that this term can be eliminated by a suitable choice of
the directions $n$ for the Wilson lines.  This is a result 
analogous to one in ref.\cite{catsey}. Whether or not this result 
generalizes to all orders remains to be investigated and is 
very important to the construction of  NLO parton showers.

The decomposition of  ref.\cite{mcsept}  entails a specific definition for 
the collinear factors, which 
         will not necessarily  
        coincide with
        the definitions used in any current event generator.  
 The issue of deriving a showering algorithm
        that corresponds to the subtracted collinear terms is the 
        subject of current work. 
        A correct answer to this
        question will encompass soft-gluon coherence and angular ordering.  
        It will      
    involve evolution equations with respect to the direction of the 
    Wilson line, Eq.~(\ref{VIVF}), in     
     terms of which the collinear subtractions are defined.

\nonumsection{Acknowledgements}
\noindent
This talk is based on work done in collaboration with J.~Collins. 
This research is funded in part by the US Department of Energy. 

\nonumsection{References}

\end{document}